\documentclass[12pt]{iopart}
\usepackage{graphicx, epsfig}
\begin{document}

\title[Thermoelectric power of Ni$_{2+x}$Mn$_{1-x}$Ga]{Composition and temperature dependence of the thermoelectric power of Ni$_{2+x}$Mn$_{1-x}$Ga alloys}
\author{P A Bhobe\dag, J H Monteiro\ddag, J C Cascalheira\ddag, S K Mendiratta\ddag, K R Priolkar\dag\footnote[3]{Author to whom correspondence should be addressed.} and P R Sarode\dag}
 \address{\dag\ Department of Physics, Goa University, Goa, 403 206 India.}
\address{\ddag\ Department of Physics, University of Aveiro, Aveiro, Portugal} 

\ead{krp@unigoa.ac.in}
\date{\today}

\begin{abstract}
The thermoelectric power (TEP) measurements have been carried out to investigate the changes in the electronic structure associated with the intermartensitic and martensitic transitions in Ni$_{2+x}$Mn$_{1-x}$Ga (0 $\le$ x $\le$ 0.19). The samples have been characterized by a.c. magnetic susceptibility measurements. The correlation between the TEP and the microstructural changes sensitive to the increasing Ni content in Ni$_{2+x}$Mn$_{1-x}$Ga has been investigated. The changes in the density of states as reflected in the temperature variation of the TEP are consistent with the conclusions drawn from the Jahn-Teller mechanism of lattice distortion. 

\end{abstract}
\pacs{72.15.Jf; 81.30.Kf; 75.50.Cc}
\submitto{\JPCM}
\maketitle

\section{Introduction}
Ni$_2$MnGa is a vastly studied ferromagnetic shape memory alloy with martensitic phase transformation taking place at T$_M$ $\sim$ 220K while the ferromagnetic order sets in at T$_C$ = 370K. The crystal structure at high temperature is the cubic L2$_1$ and transforms into a modulated tetragonal state at low temperature \cite{Web, brow14}. An important feature of the Ni-Mn-Ga alloys is the occurrence of differently modulated intermediate structures that are observed in-between the initial and the final phase with the variation of temperature \cite{mart}. The low temperature crystal structure of the martensite is a periodic stacking sequence of the \{110\}$_P$ planes along the $<$1{$\bar 1$}0$>$$_P$ direction of the initial cubic system. The phase transformation causes shuffling of atomic planes giving rise to modulations. These modulations are highly composition dependent and classification based on crystal periodicity like 5M (five layered), 7M (seven layered) or non-modulated(NM) tetragonal has been made \cite{pons}. Inelastic neutron scattering studies in the past have attributed the structural transformations to the phonon anomalies occurring in the parent phase. Here, an incomplete softening of the [$\zeta$ $\zeta$ 0]TA$_2$ phonon mode takes place at a particular wave vector $\zeta_0$(corresponding to the periodicity of the martensitic phase) with displacement along the [110] direction \cite{zhelu, str1, str2}. Such a phonon softening is believed to be due to contribution from electron-lattice coupling and nesting of the Fermi surface \cite{dug}. It is also known that alloys with low T$_M$ undergo a weak first order transformation to a micromodulated pre-martensitic phase at $\sim$ 260K upon cooling \cite{khov}. It has been well-established from earlier studies \cite{vasil} that in Ni$_{2+x}$Mn$_{1-x}$Ga, increasing Ni content raises the martensitic transformation temperature and lowers the ferromagnetic transformation temperature until they coincide at $x$ = 0.19. 

From electronic models of solids, the thermoelectric power of an alloy is known to be the function of the electron concentration, the effective mass of the electron, and the electronic scattering factor. All of these are influenced by the relative concentration of the constituent elements, lattice strain, microstructural changes, material processing and phase change. As the Ni content increases at the cost of Mn in the Ni-Mn-Ga system, there is an increase in the temperature of martensitic phase transformation. Accompanying this change is the variation in the TEP, thus making it a valued property for the phase transformation and the associated microstructural studies. Further, in metallic alloys, the value and the sign of the thermoelectric power coefficient depends on the features of the electronic bands in the vicinity of the Fermi level. Thus, TEP measurements in Ni$_{2+x}$Mn$_{1-x}$Ga should be very interesting as they would reflect the changes in the band structure occurring due to structural transformation when $x$ is varied. With this aim, we present the TEP measurements in Ni$_{2+x}$Mn$_{1-x}$Ga with $0 \le x \le 0.19$ and its correlation with the microstructural changes that accompany the structural transformation. 

\section{Experimental}
Polycrystalline ingots of Ni$_{2+x}$Mn$_{1-x}$Ga with x = 0, 0.05, 0.10, 0.13, 0.16 and 0.19 were prepared by arc-melting the constituent elements of 4N purity under argon atmosphere. To attain a good compositional homogeneity the ingots were remelted 4-5 times with a weight loss of approximately $\le$ 0.5 \% and subsequently annealed at the temperature of 1000K for 2-3 days in an evacuated quartz ampoule followed by quenching in cold water. The energy dispersive x-ray (EDX) analysis was carried out on all the ingots in order to determine the the precise composition. As can be seen from the data in Table \ref{tab:EDX}, the ratios of the constituent elements are found to be nearly equal to that of the nominal compositions.

\begin{figure}[h]
\centering
\epsfig{file=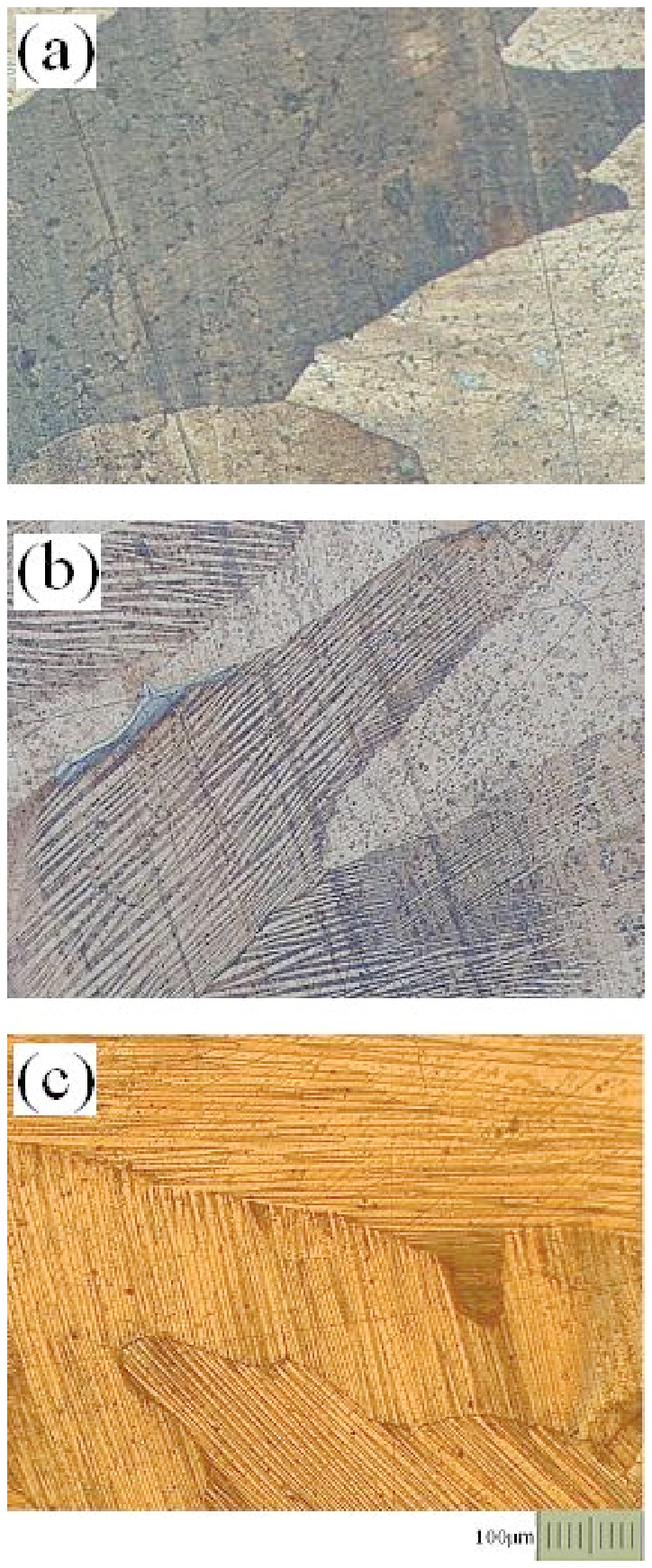, width=4cm, height=9cm}
\caption{\label{opal1}  Optical microscopy images implicating different surface morphologies exhibited by (a) $x$ = 0, (b) $x$ = 0.1 and (c) $x$ = 0.13.}
\end{figure}

The low field magnetic a.c. susceptibility was also measured in the temperature range 77K to 350K. Thermoelectric power measurements were carried out in the temperature range 100K to 350K in the warming cycle using the differential method where, the voltage difference $\Delta$V developed across the sample due to the temperature difference $\Delta$T is measured. The sample was kept between two highly polished copper plates, electrically insulated from the rest of the sample holder. Two heaters attached to these copper plates served to raise the overall temperature of the sample and to maintain a temperature gradient across the length of the sample. A copper-constantan thermocouple operating in the differential mode was employed to monitor the temperature gradient and the overall temperature of the sample was measured by a Platinum Resistance thermometer (Pt-100). By passing current through the heaters a heat pulse is generated resulting in a thermo emf ${V_{s}}$ across the sample. This voltages $V_s$ and that developed across the thermocouple  ${V_{th}}$ are measured for different temperature gradients. Thermopower of the sample is obtained by multiplying the thermopower of the thermocouple, $S_{th}$ to the slope of the straight line obtained from the plot of $V_s$ versus $V_{th}$. The microstructure of Ni$_{2+x}$Mn$_{1-x}$Ga should reflect the variation of the crystal structure as the system evolves from L2$_1$ to 5M, 7M or NM structure. Figure \ref{opal1} shows the surface morphology of $x$ = 0, 0.1, 0.13, samples which clearly reflect these microstructural changes. Martensitic transformation temperature for x = 0 being at 210K the micrograph reflects the morphology of the austenitic phase. Surface morphologies seen for $x$ = 0.1 and 0.13 are typical of a modulated structure. 

\begin{table}
\caption{\label{tab:EDX}Elemental distribution in Ni$_{2+x}$Mn$_{1-x}$Ga series as determined from EDX analysis, the average number of valence electrons per atom and the T$_M$, T$_C$  obtained from a.c. susceptibility measurements.}
\begin{indented}
\item  \begin{tabular}{@{}lcccccc}
\br
$x$ & e/a & Ni & Mn & Ga & T$_M$ & T$_C$ \\
\mr
0    & 7.5 & 50.8 & 24.6 & 24.6 & 210K & 370K\\
0.05 & 7.54 & 51.0 & 24.7 & 24.3 & 250K & 357K\\ 
0.1  & 7.58 & 52.8 & 22.5 & 24.7 & 280K & 342K\\
0.13 & 7.6 & 53.9 & 21.5 & 24.6 & 300K & 342K\\
0.16 & 7.62 & 54.4 & 21.2 & 24.4 & 315K & 337K\\
0.19 & 7.64 & 54.6 & 20.8 & 24.6 & 334K & 334K\\
\br
  \end{tabular}
\end{indented}
\end{table}

\section{Results}
 \subsection{a.c. susceptibility}
\begin{figure}[h]
\centering
\epsfig{file=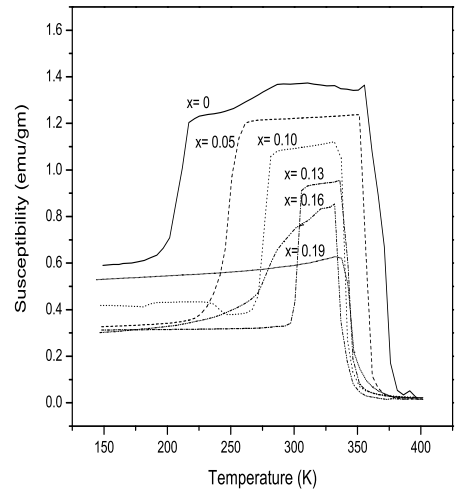, width=7cm, height=8cm}
\caption{\label{susc} a.c. susceptibility as a function of temperature for Ni$_{2+x}$Mn$_{1-x}$Ga. The martensitic and ferromagnetic phase change are marked by sudden step like features at respective T$_M$ and T$_C$ for all the alloys. As small jump in the $x$ = 0.1 at 250K indicates the inter-martensitic transformation.}
\end{figure}

Magnetic properties for the series were studied by carrying out the low field a.c. magnetic susceptibility ($\chi$) measurements as a function of temperature in the warm-up cycle and the plots are shown in figure \ref{susc}. The abrupt step-like features in $\chi$ are observed at phase transformation temperatures in all the samples. The onset of ferromagnetic transition is marked by a sharp rise in $\chi$ at T$_C$ followed by a sharp fall at the martensitic transformation temperature T$_M$. As the Ni content increases across the series, the two transitions approach each other until they merge for Ni$_{2.19}$Mn$_{0.81}$Ga at a temperature $\sim$330K. These observations are in agreement with those reported in Ref.\cite{vasil, khov1}. In addition to the features characteristic of the ferromagnetic and martensitic transformation, another feature is observed in $x$ = 0.10 at around T = 250K. This feature occurs when the sample is in the martensitic phase (T$_M$ = 285K) hence can be interpreted to be due to intermartensitic transition from one crystallographically modulated structure to the other \cite{soz}. The T$_C$ and T$_M$ for all samples as observed from the present susceptibility measurements are presented in the table \ref{tab:EDX}.

\subsection{Thermopower(TEP)}
\begin{figure}[h]
\centering
\epsfig{file=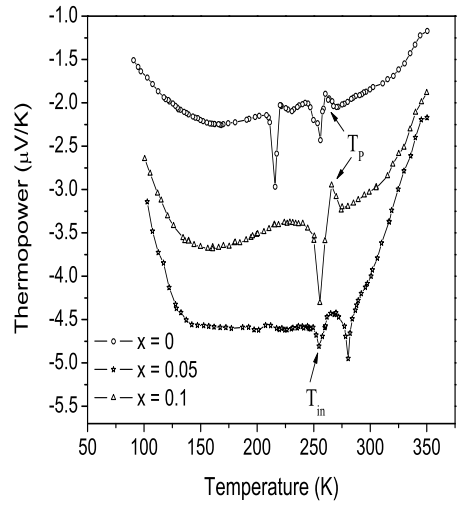, width=7cm, height=8cm}
\caption{\label{TEP-I} Thermopower as a function of temperature for Ni$_{2+x}$Mn$_{1-x}$Ga with x = 0, 0.05, 0.1. The line connecting the points is guide to the eye. The features marked as T$_P$ and T$_{in}$ indicate the pre-martensitic and intermartensitic transitions respectively.}
\end{figure}

\begin{figure}[h]
\centering
\epsfig{file=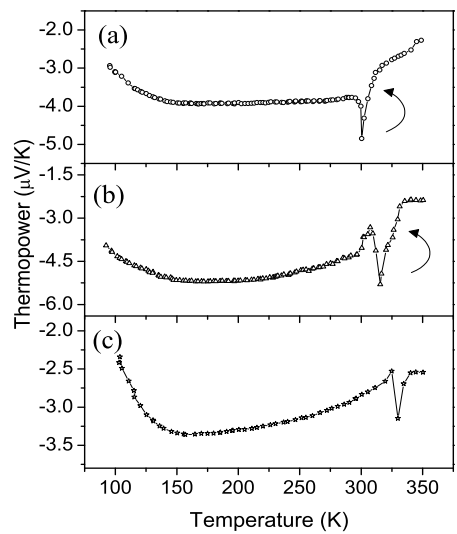, width=7cm, height=8cm}
\caption{\label{TEP-II} Thermopower as a function of temperature for Ni$_{2+x}$Mn$_{1-x}$Ga with (a)$x$ = 0.13, (b)$x$ = 0.16 and (c)$x$ = 0.19. The line connecting the points is guide to the eye. Arrows indicate the steep rise in TEP due to magnetic scattering.}
\end{figure}

Phase transformation involves a change in the free energy and TEP being sensitive to such changes is expected to show signatures of martensitic as well as magnetic transition. Thermoelectric power of Ni$_{2+x}$Mn$_{1-x}$Ga series was measured in the range 100K to 350K during warm-up and the data for all the samples is presented in figure \ref{TEP-I} and \ref{TEP-II}. It has been shown in the past that the TEP for shape memory alloys show signatures corresponding to martensitic transition \cite{lee, krp, kuo}. Indeed, as the sample is warmed a dip in TEP is obtained and can be associated with the strong phonon softening that is known to occur near the martensitic transition.On further warming a change in slope of TEP is seen due to magnetic scattering of the charge carriers as T$_C$ is approached. Thereafter, TEP attains almost a constant value that changes only slightly with temperature. This behaviour can be more clearly seen in $x \ge 0.13$. With prior knowledge of T$_M$ and T$_C$ for all the samples obtained from the magnetic susceptibility measurements, the anomalies seen in TEP data could be identified. As seen from figure \ref{TEP-I}, for $x$ = 0 the first dip at $\sim$ 210K is associated with its martensitic transformation. There is an additional anomaly occurring at $\sim$ 260K marked as T$_P$ in the figure. This feature is less intense to that occurring at T$_M$. A pre-martensitic transition is known to occur in Ni$_2$MnGa and its substitutional derivates that have a martensitic transformation temperature below 260K \cite{khov}. Hence, the anomaly marked as T$_P$ in figure \ref{TEP-I} is attributed to such pre-martensitic transition. For $x$ = 0.05, a broad spike -like feature is obtained in TEP instead of a narrow dip. As the difference in T$_M$ (=250K) and T$_P$ (=260K) is about 10K and the signatures due to these two anomalies are not well resolved in the TEP data. Two dips are also seen in $x$ = 0.1 (refer figure \ref{TEP-I}). However, martensitic transformation occurs at 280K in this sample. Hence the smaller dip at 250K is an additional feature occurring in the martensitic regime. It may be noted that the susceptibility measurements for $x$ = 0.1 also showed a signature at 250K. Thus this anomaly can be related to inter-martensitic transition. The TEP data for $x$ = 0.13, 0.16 are presented in figure \ref{TEP-II} and show huge dips in TEP at T$_M$ followed by a steep rise indicative of magnetic scattering (shown by curved arrows in figure \ref{TEP-II}). The huge dips in these samples is expected due to stronger phonon softening than in the samples with lower martensitic transition temperature \cite{str2}. For the $x$ = 0.19, martensitic and ferromagnetic transformations are known to occur at the same temperature. The TEP data for this sample indeed reflects this scenario as can be seen from figure \ref{TEP-II}(c).      

\begin{figure}[h]
\centering
\epsfig{file=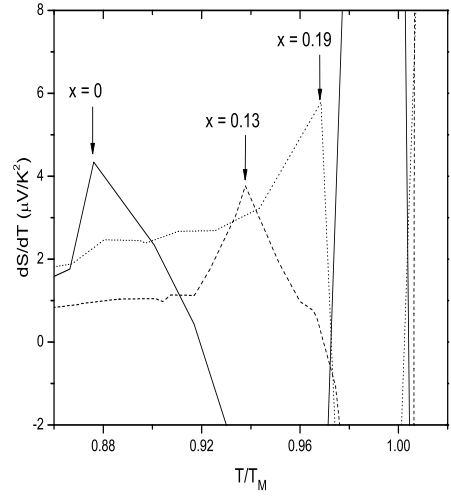, width=7cm, height=8cm}
\caption{\label{norm-tep} Temperature derivative of thermopower versus normalized temperature in Ni$_{2+x}$Mn$_{1-x}$Ga for representative concentrations. With increasing Ni, a peak is seen to move towards higher value of T/T$_M$ that can be associated with changes in DOS.}
\end{figure}

In addition to the above, another important feature is observed in TEP with the increasing Ni content. Figure \ref{norm-tep} presents the derivative of TEP as a function of normalized temperature (T/ T$_M$) for three representative samples. In the range 0.86 $\le$ T/T$_M \le$ 1.02 there lies an inflection point that shifts from 0.88 to 0.98 in going from Ni$_2$MnGa to Ni$_{2.19}$Mn$_{0.81}$Ga.

\section{Discussion}
The martensitic transformation temperatures obtained from the a.c. susceptibility measurements agree with the literature values and confirm that the Ni$_{2+x}$Mn$_{1-x}$Ga with 0 $\le x \le$ 0.19 compositions are close to nominal. In addition, we observe a intermartensitic transition in the $x$ = 0.1 sample at around T = 250K in the susceptibility data. These observations are further supported by the anomalies seen at the similar temperatures in the TEP measurements. 

Thermoelectric power data brings out many interesting features of the Ni$_{2+x}$Mn$_{1-x}$Ga series. Firstly, with increasing Ni replacing Mn atoms, the electron concentration increases and this is best reflected in the austenitic phase of all samples. In this region, a systematic increase in the magnitude of TEP is seen with increasing Ni content. This can be attributed to the electronic filling in the conduction band that takes place due to the electron transfer from nearly full 3d band of Ni to more than half filled 3d band of Mn. Further, TEP shows variations with changing concentration in the martensitic phase as well. For a typical ferromagnetic Heusler alloy, TEP goes through a minima at $\sim$ 0.4T$_C$ at low temperature \cite{camp}. A minima in TEP with a smooth parabolic type of variation is clearly seen for $x$ = 0, 0.05 samples in the low temperature region. While for $x$ = 0.1 onwards, TEP does go through a minima but thereafter, does not show a typical T$^2$ type of dependence. A reduced $|S|$ in samples with $x \ge$ 0.1 implies that the conduction electrons are getting scattered strongly after entering the martensitic phase. Such a change occurs probably as a result of different nesting vectors formed due to differently modulated crystal structures in the martensitic phase. Thus the conduction electrons get condensed differently in these changing nesting vectors. 

Another feature in TEP across the changing $x$ is the overall behaviour of TEP for $x$ = 0.19. As seen from figure \ref{TEP-I} and \ref{TEP-II} the overall magnitude of TEP in the entire temperature range increases systematically on going from $x$ = 0 to 0.16, while, $x$ = 0.19 has lower value of TEP. This behavior of 0.19 to stand out in relation to others in the sequence is perhaps the consequence of its crystal structure. As stated earlier, the crystal structure of the martensitic phase of the Ni-Mn-Ga alloys depends on composition and temperature. Studies by Lanska {\it et.al} \cite{lan} brings out a correlation between the tetragonality ($c/a$) of the martensitic structure, T$_M$ and average number of valence electrons per atom (e/a). It is shown that in the e/a range of 7.61 to 7.72, different crystal structures viz. 5M, 7M or non-modulated can be obtained upon undergoing martensitic transformation. Particularly, 5M or 7M transformations occur in samples with T$_M$ below T$_C$. In the present study, the e/a values for samples range from 7.5 to 7.64 (refer table \ref{tab:EDX}) with $x \ge$ 0.13 lying in the range 7.6 to 7.64. Also, the martensitic transition occurs in the ferromagnetic state expect for $x$ = 0.19. Thus in the martensitic region, the crystal structure for $x$ = 0.13, 0.16 is modulated and for $x$ = 0.19 it could be non-modulated \cite{pons,lan}. Also, as per the band model, there is a change in the width of the energy bands due to deformation of crystal structure and the degree of overlap of the associated orbitals changes. The fact that $x$ = 0.19 has a non-modulated structure while $x$ = 0.13, 0.16 does contain modulation, leads to deviation in the overall behaviour of TEP in Ni$_{2.19}$Mn$_{0.81}$Ga from the rest of the series.

Using the simple semi-classical result for thermal diffusion in metals, the total TEP for a ferromagnetic system can be expressed as,
\[ S = S_m + S_s\]
where $S_m$ is the magnetic contribution and \[S_s = -\frac{1}{e \sigma T} \int (\epsilon - \mu)\frac{\partial f_0}{\partial \epsilon}\sigma(\epsilon)d\epsilon\]
where $f_0$ is the Fermi-Dirac distribution function,~ $\mu$ is the chemical potential and \[ \sigma = \int \sigma(\epsilon)\frac{\partial f_0}{\partial \epsilon}d\epsilon\]
It can be clearly seen that TEP is dependent on density of states (DOS) near the Fermi level. The above result has been used to explain the TEP of shape memory alloy, NiTi assuming a DOS with a peak near Fermi level \cite{lee}.

The alloys in the present study are in the ferromagnetic state and hence the magnetic scattering can be assumed to be similar for all the samples. Thus change in TEP with respect to the concentration of alloys can be related only to changes in DOS at the Fermi level. Further, the band structure of Ni$_2$MnGa has been calculated and the composition of bands that are active at the Fermi surface have been identified \cite{fuj,zay}. According to these calculation there is a peak in DOS just below the Fermi level comprising of minority spin Ni $e_g$ band. A peak is indeed seen in the temperature derivative of TEP near the martensitic transition temperature. This is presented in figure \ref{norm-tep} where temperature derivative of TEP with respect to normalised temperature (T/T$_M$) is plotted for three representative samples. Moreover, this peak shifts from 0.88 to 0.98 in going from Ni$_2$MnGa to Ni$_{2.19}$Mn$_{0.81}$Ga. Such a shift is an indication of changes occuring in the DOS within the martensitic phase. Upon martensitic transformation, a structural change from cubic to orthorhombic leads to lifting of degeneracy of 3d levels. It is the splitting of energy sub-bands which are degenerate in the cubic phase that enables the electrons to redistribute themselves in order to lower the free energy. This is the well known band Jahn-Teller mechanism and has been observed in Ni$_2$MnGa using polarised neutron scattering \cite{brow11}. For a martensitic transition to occur it is required that the peak in the DOS should have some asymmetry, whereby it has more weighting towards lower energies and that the Fermi level is situated very close to the peak. From the present TEP data, the shift of the inflection point from 0.88 to 0.98 with increasing Ni content can be attributed to the movement of the Ni $e_g$ band towards Fermi level thereby leading to increase in T$_M$. 

\section{Conclusion}
Thermoelectric power measurements on the Ni$_{2+x}$Mn$_{1-x}$Ga with 0 $\le x \le$ 0.19 exhibit the signatures not only for the martensitic transitions but also the premartensitic and intermartensitic transitions in some of the compositions. Thermopower reflects the changes in the density of states consistent with the changes in the electronic structure due to band Jahn Teller effect. The increase in martensitic transformation temperatures with increasing Ni content can be related to the shifting of a peak in DOS at Fermi level closer to $E_F$. The change in behaviour of TEP from that expected for a typical Heusler alloy, in the martensitic region can be related to the differently modulated crystal structures formed upon martensitic transformation.   

\ackn
P A Bhobe acknowledges Council for Scientific and Industrial Research, New Delhi for the Senior Research Fellowship. A part of this work was carried out under DST-GRICES funded project. Financial assistance by DST, India and GRICES, Portugal and local hospitality by University of Aveiro to K R Priolkar is gratefully acknowledged.

\newpage

\end{document}